# Feasibility Study of Neural ODE and DAE Modules for Power System Dynamic Component Modeling

Tannan Xiao, *IEEE Member*, Ying Chen, *IEEE Member*, Shaowei Huang, *IEEE Member*, Tirui He, and Huizhe Guan

*Abstract*—In the context of high penetration of renewables, the need to build dynamic models of power system components based on accessible measurement data has become urgent. To address this challenge, firstly, a neural ordinary differential equations (ODE) module and a neural differential-algebraic equations (DAE) module are proposed to form a data-driven modeling framework that accurately captures components' dynamic characteristics and flexibly adapts to various interface settings. Secondly, analytical models and data-driven models learned by the neural ODE and DAE modules are integrated together and simulated simultaneously using unified transient stability simulation methods. Finally, the neural ODE and DAE modules are implemented with Python and made public on GitHub. Using the portal measurements, three simple but representative cases of excitation controller modeling, photovoltaic power plant modeling, and equivalent load modeling of a regional power network are carried out in the IEEE-39 system and 2383wp system. Neural dynamic model-integrated simulations are compared with the original model-based ones to verify the feasibility and potentiality of the proposed neural ODE and DAE modules.

## I. Introduction

### A. Motivations

Ensuring the stable operation of power systems is a crucial task, which relies heavily on accurate system dynamic simulations [1]. However, after adopting massive renewables, power systems have become far more complicated, which raises new technical challenges in this area.

First, without enough prior knowledge, it is hard to develop dynamic models of electrical and control components as well as equivalent models of subsystems deductively and precisely [2]. Impacts of hidden dynamics, e.g., environmental changes, on modern power systems are strengthened as more renewables are integrated into all voltage levels. Lacking knowledge of such hidden dynamics poses difficulties in deriving equivalent models of load areas [3] and renewable power plants [4], etc. Moreover, for privacy considerations, vendors and users may not fully share the knowledge of controls and protections of their power devices [5]. From the point of view of system operators, these devices and controls are black-boxes weakening the accurate understanding of system dynamics.

Second, there are few power system simulation tools for integrating analytical and data-driven models seamlessly [6]. Utilizing the big data of real-time measurements and simulation results, various data-driven surrogate models have been developed in this decade [7], which were used to predict local and system-level dynamics of power systems. However, these data-driven models were seldom deployed directly in traditional simulation programs to perform numerical integration with analytical models of power components. Such an absence results not only from incompatible forms of data-driven models but also lacking numerical algorithms integrating both analytical and data-driven models together.

To solve the above challenging problems, we explore the feasibility of building power system dynamic models in a data-driven manner based on neural ordinary differential equations (ODE). Moreover, unified simulation methods are also introduced to integrate power component models both in analytical and neural network forms. Neural dynamic models of controllers, photovoltaic power plants, and subsystems are developed for illustration and tests. Comprehensive case studies validate the efficacy of the proposed method.

### B. Related Works

Dynamic modeling methods can be divided into three types, namely, the analytical approach based on knowledge, the data-driven approach based on measurement data, and the hybrid approach combining the former two [8]. There are mainly three lines of studies that inspire the data-driven modeling method proposed in this paper, namely, neural ODE, physics-informed neural networks (PINN), and the Koopman operator. Here we provide a brief introduction to these topics.

The idea of modern neural ODE was proposed by Chen *et al* in 2018 [9]. As for the analytical modeling approach and the data-driven modeling approach, it was stated in [10] that neural ODE offers a best-of-both-worlds approach. While learning a global approximation of the derivative functions with easily trainable neural networks, neural ODE also keeps the classical numerical integration structure, which is a very important prior knowledge and makes neural ODE highly adaptive to scientific computations and industrial applications. This idea quickly received worldwide attention and inspired many other studies such as neural partial differential equations (PDE) [11], the Hamilton neural networks [12], the Lagrangian neural networks [13], etc. It should be noted that the basic idea of neural ODE

This paper is first submitted to IEEE. This work was supported in part by the National Natural Science Foundation of China (NSFC) under the Project No. 52107104 and Project No. 51861135312, (*Corresponding Author: Ying Chen*).

All the authors are with the Department of Electrical Engineering, Tsinghua University, Beijing, 100084, China (Email: eexiaoxh@gmail.com, chen_ying@tsinghua.edu.cn, huangsw@tsinghua.edu.cn, hetirui@qq.com, mark.mghz@gmail.com).

can be found in older literature, e.g., in the research field of power systems, a similar idea can be traced back to 2003 [14] for equivalent load modeling, which used a bottleneck neural network to extract state variables from algebraic variables and embedded a neural network to learn the time derivatives of the extracted state variables. Another example is [15], which adopted the idea to approximate differential-algebraic equations (DAE) for the equivalent modeling of a regional power network. Recently in 2022, the learning theory of neural ODE was introduced and a neural ODE-based equivalent modeling of networked microgrids was built for reachability analysis [16]. Currently, the learning theory of power system DAE approximation is not clarified. Meanwhile, the trained neural dynamic models are seldom integrated directly into traditional power system simulators to perform dynamic simulations with analytical models at the same time.

Different from neural ODE, the idea of PINN is to directly learn an approximation of the solutions of numerical integration methods, which was proposed to solve PDEs at first [17]. The output of the neural networks contains the solution of the current integration step and the hidden vector for the next integration step. The PINN works like a recurrent neural network and no longer needs to fit in a numerical ODE solver. A PINN for DAE solutions called DAE-PINN was proposed in [18] to directly output the solutions of required state variables and algebraic variables of the IEEE-9 power system, which means that a system-level surrogate is learned with limited observations.

The idea of the Koopman operator was proposed in 1931 [19] and 1932 [20]. The Koopman operator is a linear but infinite-dimensional operator that governs the evolution of functions of possible measurements of the system state. With the recent development of machine learning theory, the Koopman operator can be learned using an autoencoder structure and has become a leading candidate for the global linear representation of non-linear systems [21]. An example of the Koopman operator theory in power systems is to identify system modes in a data-driven manner using ambient system measurements [22].

### C. Contributions

The contributions of this paper are as follows.

1) A neural ODE module with external inputs (neural ODE-E module) and a neural DAE module are proposed to build accurate data-driven dynamic models for power system components based on accessible measurement data.

2) Neural model-integrated transient stability simulation methods are presented to integrate data-driven models trained by the proposed neural modules with analytical models and perform simulations simultaneously without affecting the accuracy and convergence of solutions.

3) The proposed neural modules are implemented with Python and the source code is made public on GitHub[1]. Using the portal measurements, three simple but representative cases of excitation controller modeling, photovoltaic power plant modeling, and equivalent load modeling of a regional power network are carried out in the IEEE-39 system and 2383wp system to validate the feasibility of the proposed neural modules and neural dynamic model-integrated simulations. Test results indicate that the proposed neural modules can build accurate dynamic models for complex components based on a dataset of portal measurements that only contains stable samples with a limited proportion of large-disturbance contingencies.

---

[1] https://github.com/xxh0523/Py_PSNODE

### D. Paper Organization

The remainder of the papers is as follows. Section II introduces the problem formulation of power system dynamic component modeling. The neural ODE-E module and neural DAE module are illustrated in Section III. In section IV, how to integrate the neural models into transient stability simulations is demonstrated. Numerical tests are designed in Section V and test results are shown in Section VI. Discussions on the proposed neural modules are carried out in Section VII. Conclusions are drawn in Section VIII.

## II. PROBLEM FORMULATION

### A. Power System Component Modeling and Simulations

In power system transient stability simulations, the dynamic components can be divided into two types according to whether the component produces injection currents into the power network, as shown in Fig. 1. For the components that do not produce injection currents (hereafter referred to as *controllers*), the dynamics can be formulated as:

$$\dot{\mathbf{x}} = \mathbf{f}(\mathbf{x}, \mathbf{z}) \tag{1}$$

where $\mathbf{x}$ is the state vector of the component whose time derivatives are equal to $\mathbf{f}(\mathbf{x}, \mathbf{z})$, and $\mathbf{z}$ is the vector of required external input variables. Taking an exciter as an example, in addition to the state variables, the nodal voltage of the generator-connected bus is needed for calculating derivatives, i.e., the nodal voltage belongs to $\mathbf{z}$.

For the components that produce injection currents into the power network (hereafter referred to as *power devices*), the dynamics can be formulated as:

$$\dot{\mathbf{x}} = \mathbf{f}(\mathbf{x}, \mathbf{i}, \mathbf{v}, \mathbf{z}) \tag{2}$$

$$\mathbf{i} = \mathbf{g}(\mathbf{x}, \mathbf{v}, \mathbf{z}) \tag{3}$$

where $\mathbf{i}$ is the injection current, $\mathbf{v}$ is the nodal voltage of the component-connected bus, and $\mathbf{g}$ is the function of injection current calculation. Taking a synchronous machine as an example, the inner electric potentials, rotor angle, rotor speed, and state variables of controllers belong to $\mathbf{x}$, and the active power and the reactive power at the starting instant and at the current step belong to $\mathbf{z}$.

Accordingly, in transient stability simulations, the overall dynamics of the power system are modeled as a group of high-dimensional non-linear DAEs including ODEs (4) for dynamic devices and algebraic equations (AE) (5) for the power network:

$$\dot{\mathbf{x}} = \mathbf{f}(\mathbf{x}, \mathbf{V}) \tag{4}$$

$$\mathbf{G}(\mathbf{x}, \mathbf{V}) = \mathbf{Y}\mathbf{V} - \mathbf{I}(\mathbf{x}, \mathbf{V}) = \mathbf{0} \tag{5}$$

where $\mathbf{x}$ is the state vector of the system, whose time derivatives are equal to $\mathbf{f}(\mathbf{x}, \mathbf{V})$, $\mathbf{V}$ is the bus voltage vector, $\mathbf{I}$ is the injection current vector, $\mathbf{Y}$ is the admittance matrix, and $\mathbf{G}$ represents the whole network equations. The system DAEs can be solved with the simultaneous approach or the alternating approach [23]. It is obvious that the ODE models shown in (1) and DAE models shown in (2) and (3) can be easily integrated into the system DAEs shown in (4) and (5).

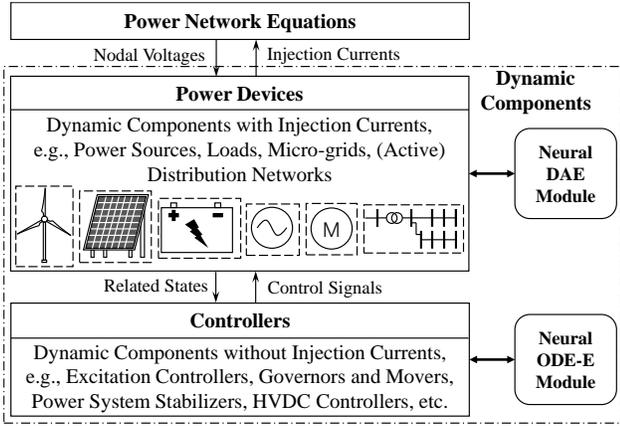

Fig. 1. Overview of power system dynamic component modeling and proposed neural modules.

### B. Data-driven Dynamic Component Modeling

This paper focuses on building dynamic component models based on accessible measurement data and integrating the neural models into power system time-domain simulators. The overview of the proposed method is shown in Fig. 1.

Typically, in practical power systems, the portal measurements of a dynamic component, i.e., the input and output of controllers and the nodal voltages and injection currents of power devices, are available, whereas the inner variables and dynamics of the component are commonly not observable. These accessible measurements are assembled as:

$$\mathcal{S} = \{\hat{\mathbf{x}}, \hat{\mathbf{i}}, \hat{\mathbf{v}}, \hat{\mathbf{z}}\} \qquad (6)$$

where $\mathcal{S}$ denotes the dataset, and $\hat{\mathbf{x}}$, $\hat{\mathbf{i}}$, $\hat{\mathbf{v}}$, and $\hat{\mathbf{z}}$ represent the measured values of $\mathbf{x}$, $\mathbf{i}$, $\mathbf{v}$, and $\mathbf{z}$, respectively. Notably, $\hat{\mathbf{x}}$ only contains measurable state variables, e.g., the output of a controller, and is set to empty if no state variable is available.

Based on $\mathcal{S}$, the neural ODE-E module introduced in section III.B is used to build data-driven dynamic models for controllers, while the neural DAE module presented in section III.C is used to build data-driven dynamic models for power devices.

The data-driven model derived from $\mathcal{S}$ needs to adapt to the mature framework of power system dynamic simulations and be simulated simultaneously with analytical models without seriously affecting the accuracy and convergence of solutions, which is introduced in section IV.

### III. NEURAL ODE-E MODULE AND NEURAL DAE MODULE FOR POWER SYSTEM DYNAMIC COMPONENT MODELING

### A. General Framework of Neural ODE

The structure of the original neural ODE module, as shown in Fig. 2(a), was first proposed by Chen *et al* [9]. A neural ODE block consisting of neural networks is used to formulate a parameterized derivative function $\Psi(\mathbf{x};\theta)$, as shown in (7):

$$\Psi(\mathbf{x};\theta) \doteq \dot{\mathbf{x}} = \mathbf{f}(\mathbf{x}) \qquad (7)$$

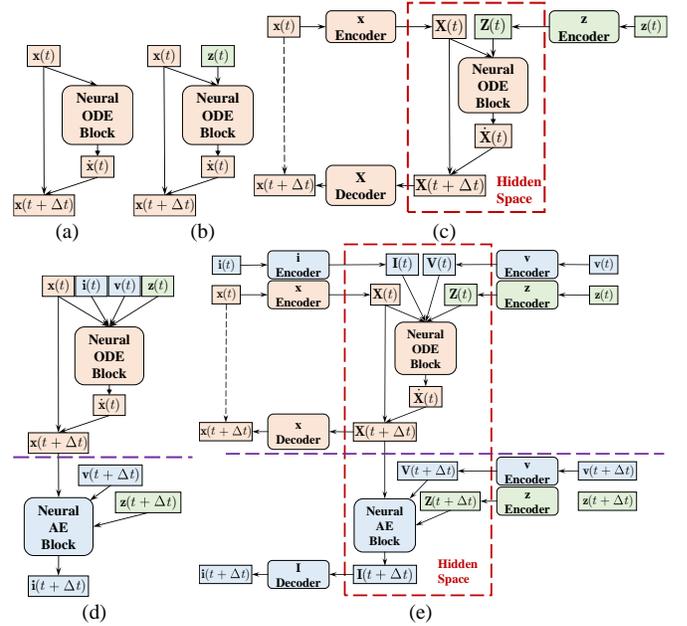

Fig. 2. Structures of (a) original neural ODE module, (b) neural ODE-E module, (c) autoencoder-based neural ODE-E module, (d) neural DAE module, and (e) autoencoder-based neural DAE module.

where $\theta$ denotes parameters of the neural ODE block. A simple example is shown in (8) when a multi-layer perceptron (MLP) with one hidden layer is used in the neural ODE block:

$$\dot{\mathbf{x}} = \sigma\left[\mathbf{W}_2(\mathbf{W}_1\mathbf{x}+\mathbf{b}_1)+\mathbf{b}_2\right], \theta = \{\mathbf{W}_1,\mathbf{b}_1,\mathbf{W}_2,\mathbf{b}_2\} \qquad (8)$$

where $\mathbf{W}$ and $\mathbf{b}$ denote the parameters of the MLP, and $\sigma$ is the activation function.

Given the initial values $\mathbf{x}(0)$, the integration step $\Delta t$, the integration time $T$, and the derivative function (7), a series of $\mathbf{x}$ can be easily calculated with a numerical integration method. Both the explicit and implicit integration methods can be used to solve neural ODE. For the convenience of demonstration, all the modules displayed in Fig. 2 use the Euler method as an example and the stepwise integration is as follows:

$$\mathbf{x}(t+\Delta t) = \mathbf{x}(t) + \Delta t \cdot \mathbf{f}(\mathbf{x}(t)) \qquad (9)$$

The neural ODE block can be trained with a dataset of sampled curves of $\mathbf{x}$. In (10), a loss function $\mathcal{L}$, e.g., the mean square error (MSE) between predicted curves and ground-truth curves, is defined to optimize $\theta$:

$$\theta = \arg\min_{\theta}\sum_{i=1}^{N}\mathcal{L}(\mathbf{x}_i,\hat{\mathbf{x}}_i;\theta)$$
$$\text{s.t.} \quad \dot{\mathbf{x}}_i = \Psi(\mathbf{x};\theta), \mathbf{x}_i(0) = \hat{\mathbf{x}}_i(0), \hat{\mathbf{x}}_i \in \mathcal{S}, t \in [0,T] \qquad (10)$$

where $N$ denotes the number of samples in the training set, $\mathbf{x}$ denotes the series of state variables predicted using the neural ODE module, $\mathcal{S}$ denotes the sample set, and $\hat{\mathbf{x}}$ denotes the ground-truth series. With the initial value $\hat{\mathbf{x}}_i(0)$, $\mathbf{x}_i$ can be obtained by numerical integration.

To optimize $\theta$, the adjoint sensitivity method [24], which is also introduced in [9] and [16], can be adopted. The gradients with respect to $\theta$ are calculated as shown in (11):

$$\nabla_{\theta}\mathcal{L} = \frac{\partial \mathcal{L}}{\partial \theta} = \int_0^T \frac{\partial \mathcal{H}(t)}{\partial \theta}dt = \int_0^T \frac{\partial\{\Psi[\mathbf{x}(t);\theta]\}^T}{\partial \theta}\lambda(t)dt \qquad (11)$$

where $\mathcal{H}(t) = \mathcal{H}[\mathbf{x}(t);\theta,\lambda(t)]$ is the Hamiltonian function [25]

defined in (12) and $\lambda$ denotes the Lagrange multiplier that obeys (13)-(14):

$$\mathcal{H}(\mathbf{x};\boldsymbol{\theta},\boldsymbol{\lambda}) = \boldsymbol{\lambda}^{\mathrm{T}}\boldsymbol{\Psi}(\mathbf{x};\boldsymbol{\theta}) \tag{12}$$

$$\dot{\boldsymbol{\lambda}} = -\frac{\partial \mathcal{H}}{\partial \mathbf{x}} = -\frac{\partial [\boldsymbol{\Psi}(\mathbf{x};\boldsymbol{\theta})]^{\mathrm{T}}}{\partial \mathbf{x}}\boldsymbol{\lambda} \tag{13}$$

$$\boldsymbol{\lambda}(T) = \frac{\partial \mathcal{L}}{\partial \mathbf{x}(T)}, \quad \boldsymbol{\lambda}(t) = \frac{\partial \mathcal{L}}{\partial \mathbf{x}(t)} + \int_T^t \dot{\boldsymbol{\lambda}} dt \tag{14}$$

The proof of the gradient formula can be found in Appendix. A. Like the forward integration, a numerical integration method can be used to solve the backward integration shown in (14) using a negative integration step $-\Delta t$. The parameters can be easily optimized after the gradients $\nabla_{\boldsymbol{\theta}}\mathcal{L}$ are obtained, e.g., by applying the stochastic gradient descent (SGD) method or the Adam optimizer [26].

*B. Neural ODE-E Module*

Based on the general framework of neural ODE, for the components shown in (1) that do not produce injection currents into the power network, the dynamics of the component and the neural derivative function can be formulated as (15):

$$\boldsymbol{\Psi}(\mathbf{x},\mathbf{z};\boldsymbol{\theta}) \doteq \dot{\mathbf{x}} = \mathbf{f}(\mathbf{x},\mathbf{z}) \tag{15}$$

Accordingly, the neural ODE-E module is shown in Fig. 2(b). By substituting (15) into (10), the overall optimization of the neural ODE-E module is formulated as:

$$\boldsymbol{\theta} = \arg\min_{\boldsymbol{\theta}} \sum_{i=1}^{N} \mathcal{L}(\mathbf{x}_i, \hat{\mathbf{x}}_i)$$
$$\text{s.t.} \quad \dot{\mathbf{x}}_i = \boldsymbol{\Psi}(\mathbf{x}_i, \hat{\mathbf{z}}_i;\boldsymbol{\theta}), \mathbf{x}_i(0) = \hat{\mathbf{x}}_i(0), \{\hat{\mathbf{x}}_i, \hat{\mathbf{z}}_i\} \in \mathcal{S}, t \in [0,T] \tag{16}$$

where $\hat{\mathbf{z}}$ denotes the ground-truth values of $\mathbf{z}$. After applying the adjoint sensitivity method, parameters $\boldsymbol{\theta}$ can be optimized as in (12)-(14) by changing $\boldsymbol{\Psi}(\mathbf{x};\boldsymbol{\theta})$ to $\boldsymbol{\Psi}(\mathbf{x},\hat{\mathbf{z}};\boldsymbol{\theta})$.

*C. Neural DAE Module*

For the components shown in (2) and (3) that produce injection currents into the power network, a more complex module illustrated in Fig. 2. (d) is designed by adding a neural AE block. The dynamics can be approximated using the neural DAE module as follows:

$$\begin{cases} \boldsymbol{\Psi}(\mathbf{x},\mathbf{i},\mathbf{v},\mathbf{z};\boldsymbol{\theta}) \doteq \dot{\mathbf{x}} = \mathbf{f}(\mathbf{x},\mathbf{i},\mathbf{v},\mathbf{z}) \\ \boldsymbol{\Phi}(\mathbf{x},\mathbf{i},\mathbf{v},\mathbf{z};\boldsymbol{\xi}) \doteq \mathbf{i} - \boldsymbol{\varphi}(\mathbf{x},\mathbf{v},\mathbf{z};\boldsymbol{\xi}) = \mathbf{0} \\ = \mathbf{G}(\mathbf{x},\mathbf{i},\mathbf{v},\mathbf{z}) \doteq \mathbf{i} - \mathbf{g}(\mathbf{x},\mathbf{v},\mathbf{z}) \end{cases} \tag{17}$$

where $\boldsymbol{\Phi}(\mathbf{x},\mathbf{i},\mathbf{v},\mathbf{z};\boldsymbol{\xi})$ denotes the neural AEs, $\boldsymbol{\varphi}(\mathbf{x},\mathbf{v},\mathbf{z};\boldsymbol{\xi})$ denotes the neural AE block, and $\boldsymbol{\xi}$ denotes the parameters of the neural AE block. Similarly, the optimization of $\boldsymbol{\theta}$ and $\boldsymbol{\xi}$ can be formulated as shown in (18). The adjoint sensitivity method can also be used to derive gradients with respect to $\boldsymbol{\theta}$ and $\boldsymbol{\xi}$, which has the same form as the trajectory sensitivity method in the research field of power systems [27].

$$\boldsymbol{\theta},\boldsymbol{\xi} = \arg\min_{\boldsymbol{\theta},\boldsymbol{\xi}} \sum_{i=1}^{N} \mathcal{L}(\mathbf{x}_i, \hat{\mathbf{x}}_i, \mathbf{i}_i, \hat{\mathbf{i}}_i; \boldsymbol{\theta},\boldsymbol{\xi})$$
$$\text{s.t.} \begin{cases} \dot{\mathbf{x}}_i = \boldsymbol{\Psi}(\mathbf{x}_i, \mathbf{i}_i, \hat{\mathbf{v}}_i, \hat{\mathbf{z}}_i;\boldsymbol{\theta}) \\ \boldsymbol{\Phi}(\mathbf{x}_i, \mathbf{i}_i, \hat{\mathbf{v}}_i, \hat{\mathbf{z}}_i;\boldsymbol{\xi}) = \mathbf{i}_i - \boldsymbol{\varphi}(\mathbf{x}_i, \hat{\mathbf{v}}_i, \hat{\mathbf{z}}_i;\boldsymbol{\xi}) = \mathbf{0} \end{cases} \tag{18}$$
$$\mathbf{x}_i(0) = \hat{\mathbf{x}}_i(0), \mathbf{i}_i(0) = \hat{\mathbf{i}}_i(0), \{\hat{\mathbf{x}}_i, \hat{\mathbf{i}}_i, \hat{\mathbf{v}}_i, \hat{\mathbf{z}}_i\} \in \mathcal{S}, t \in [0,T]$$

The gradients can be calculated as shown in (19) and (20):

$$\nabla_{\boldsymbol{\theta}}\mathcal{L} = \frac{\partial \mathcal{L}}{\partial \boldsymbol{\theta}} = \int_0^T \frac{\partial \mathcal{H}(t)}{\partial \boldsymbol{\theta}} dt = \int_0^T \frac{\partial [\boldsymbol{\Psi}(t)]^{\mathrm{T}}}{\partial \boldsymbol{\theta}} \boldsymbol{\lambda}(t) dt \tag{19}$$

$$\nabla_{\boldsymbol{\xi}}\mathcal{L} = \frac{\partial \mathcal{L}}{\partial \boldsymbol{\xi}} = \int_0^T \frac{\partial \mathcal{H}(t)}{\partial \boldsymbol{\xi}} dt = \int_0^T \frac{\partial [\boldsymbol{\Phi}(t)]^{\mathrm{T}}}{\partial \boldsymbol{\xi}} \boldsymbol{\beta}(t) dt \tag{20}$$

where $\boldsymbol{\Psi}(t)$ denotes $\boldsymbol{\Psi}(\mathbf{x}(t),\mathbf{i}(t),\hat{\mathbf{v}}(t),\hat{\mathbf{z}}(t);\boldsymbol{\theta})$, $\boldsymbol{\Phi}(t)$ denotes $\boldsymbol{\Phi}(\mathbf{x}(t),\mathbf{i}(t),\hat{\mathbf{v}}(t),\hat{\mathbf{z}}(t);\boldsymbol{\xi})$, $\boldsymbol{\varphi}(t)$ denotes $\boldsymbol{\varphi}(\mathbf{x}(t),\hat{\mathbf{v}}(t),\hat{\mathbf{z}}(t);\boldsymbol{\xi})$, $\mathcal{H}$ denotes the Hamiltonian function defined in (21), $\mathcal{H}(t)$ denotes $\mathcal{H}(\mathbf{x}(t),\mathbf{i}(t),\hat{\mathbf{v}}(t),\hat{\mathbf{z}}(t);\boldsymbol{\theta},\boldsymbol{\xi},\boldsymbol{\lambda}(t),\boldsymbol{\beta}(t))$, and $\boldsymbol{\lambda}$ and $\boldsymbol{\beta}$ denote the Lagrange multipliers that obey (22)-(23):

$$\mathcal{H}(\mathbf{x},\mathbf{i},\hat{\mathbf{v}},\hat{\mathbf{z}};\boldsymbol{\theta},\boldsymbol{\xi},\boldsymbol{\lambda},\boldsymbol{\beta})$$
$$= \boldsymbol{\lambda}^{\mathrm{T}}\boldsymbol{\Psi}(\mathbf{x},\mathbf{i},\hat{\mathbf{v}},\hat{\mathbf{z}};\boldsymbol{\theta}) + \boldsymbol{\beta}^{\mathrm{T}}\boldsymbol{\Phi}(\mathbf{x},\mathbf{i},\hat{\mathbf{v}},\hat{\mathbf{z}};\boldsymbol{\xi}) \tag{21}$$

$$\begin{cases} \dot{\boldsymbol{\lambda}} = -\frac{\partial \mathcal{H}}{\partial \mathbf{x}} = -\frac{\partial \boldsymbol{\Psi}^{\mathrm{T}}}{\partial \mathbf{x}}\boldsymbol{\lambda} - \frac{\partial \boldsymbol{\Phi}^{\mathrm{T}}}{\partial \mathbf{x}}\boldsymbol{\beta} \\ \mathbf{0} = \frac{\partial \mathcal{H}}{\partial \mathbf{i}} = \frac{\partial \boldsymbol{\Psi}^{\mathrm{T}}}{\partial \mathbf{i}}\boldsymbol{\lambda} + \frac{\partial \boldsymbol{\Phi}^{\mathrm{T}}}{\partial \mathbf{i}}\boldsymbol{\beta} \end{cases} \tag{22}$$

$$\begin{cases} \boldsymbol{\lambda}(T) = \frac{\partial \mathcal{L}}{\partial \mathbf{x}(T)} + \frac{\partial [\boldsymbol{\varphi}(T)]^{\mathrm{T}}}{\partial \mathbf{x}(T)}\frac{\partial \mathcal{L}}{\partial \mathbf{i}(T)} \\ \boldsymbol{\lambda}(t) = \frac{\partial \mathcal{L}}{\partial \mathbf{x}(t)} + \frac{\partial [\boldsymbol{\varphi}(t)]^{\mathrm{T}}}{\partial \mathbf{x}(t)}\frac{\partial \mathcal{L}}{\partial \mathbf{i}(t)} + \int_T^t \dot{\boldsymbol{\lambda}} dt \end{cases} \tag{23}$$

The proof of the gradient formulas can be found in Appendix. A. Similarly, backward numerical integration is required to solve $\boldsymbol{\lambda}$ and $\boldsymbol{\beta}$ before the gradients with respect to $\boldsymbol{\theta}$ and $\boldsymbol{\xi}$ can be derived.

*D. Autoencoder-based frameworks*

Frameworks of autoencoder-based neural modules are put forward in Fig. 2. (c) and Fig. 2. (e).

From a practical point of view, dimensionality change is an intuitive demand of power systems. On the one hand, real-world dynamic components are not completely observable in most cases. None or only a very limited number of state variables can be measured. The dynamics of the unobservable part can be approximated by a dimensionality-raising autoencoder based on the sampled curves of the observable part. On the other hand, in an equivalent modeling situation, there are too many state variables. In this case, a dimensionality-reduction autoencoder can be adopted to focus on the key features of the original dynamics.

From a theoretical point of view, the introduction of autoencoders can improve the flexibility and performance of the proposed neural modules. The key function of these autoencoders is to embed the original states and dynamics into a hidden space. On the one hand, a difficult problem in the original low-dimensional space can change to a simpler one after a proper dimensionality-raising autoencoding. The same idea is also used in kernel-based support vector machines. On the other hand, for complex high-dimensional systems, dimensionality-reduction autoencoders can be used to find existing low-dimensional manifolds, even to find Koopman invariant subspaces and linearize the system dynamics [21].

Therefore, in this paper, autoencoders are used to change dimensionality and improve the flexibility and performance of the neural modules, as shown in (24) and (25):

$$\begin{cases} \mathbf{X}(0) = enc_{\mathbf{x}}(\mathbf{x}(0)), \mathbf{Z}(t) = enc_{\mathbf{z}}(\mathbf{z}(t)) \\ \mathbf{x}(t) = dec_{\mathbf{x}}(\mathbf{X}(t)) \end{cases} \quad (24)$$

$$\begin{cases} \mathbf{X}(0) = enc_{\mathbf{x}}(\mathbf{x}(0)), \mathbf{I}(0) = enc_{\mathbf{i}}(\mathbf{i}(0)) \\ \mathbf{V}(t) = enc_{\mathbf{v}}(\hat{\mathbf{v}}(t)), \mathbf{Z}(t) = enc_{\mathbf{z}}(\hat{\mathbf{z}}(t)) \\ \mathbf{x}(t) = dec_{\mathbf{x}}(\mathbf{X}(t)), \mathbf{i}(t) = dec_{\mathbf{I}}(\mathbf{I}(t)) \end{cases} \quad (25)$$

where $enc$ denotes an encoder, $dec$ denotes a decoder, and $\mathbf{X}$, $\mathbf{Z}$, $\mathbf{V}$, and $\mathbf{I}$ are the state variables and algebraic variables in the hidden space. The numerical integrations take place in the hidden space. It should be noted that the encoding of $\mathbf{x}$ and $\mathbf{i}$ is only performed once to transform the initial value of $\mathbf{x}(0)$ and $\mathbf{i}(0)$ into $\mathbf{X}(0)$ and $\mathbf{I}(0)$ in the hidden space. By solving neural DAE, $\mathbf{X}(t)$ and $\mathbf{I}(t)$ are derived in the hidden space. Therefore, the decoding process is performed at every integration step to derive $\mathbf{x}(t)$ and $\mathbf{i}(t)$ in the original space. The frameworks displayed in Fig. 2. (c) and Fig. 2. (e) are adopted for ease of description and understanding. The autoencoders are simultaneously optimized with the neural blocks.

### E. Loss Function and Training Procedures

---
**Pseudo-code 1: Training Procedures of Neural Modules**

Input the total number of samples $N$, the integration step $\Delta t$, the simulation time $T$, the mini-batch size $m$, the number of training epochs $E$, and the evaluation interval $\Delta E$.

Get the sampled dataset $\mathcal{S}$ by simulations with simulation time $T$ and fixed integration step $\Delta t$ or by real-time measurements. Split the samples to form a training set and a testing set.

**for** $i = 1$ **to** $E$ **do**
  **for** $j = 1$ **to** $N/m$ **do**
    Sample a mini-batch of $m$ samples from the training dataset.
    Input the mini-batch into the neural module, perform forward integration, and get $\mathbf{x}$ and $\mathbf{i}$.
    Perform backward integration and get $\nabla_\theta \mathcal{L}$ and $\nabla_\xi \mathcal{L}$.
    Update $\theta$ and $\xi$ with the optimizer based on $\nabla_\theta \mathcal{L}$ and $\nabla_\xi \mathcal{L}$.
  **end for**
  **if** $i$ mod $\Delta E$ **then**
    Evaluate the neural module in the testing dataset.
  **end if**
**end for**

---

For the convenience of demonstration, we take $(\mathbf{x}, \mathbf{i}, \mathbf{v}, \mathbf{z})$ and the neural DAE module as the example to explain the loss function. The loss function $\mathcal{L}$ can be calculated as:

$$\mathcal{L} = \sum_{k=0}^{n} \mathcal{L}(t_k), \quad t_k \in [0, T], t_0 = 0, t_n = T \quad (26)$$

where $\mathcal{L}$ denotes $\mathcal{L}(\mathbf{x}_i(t_k), \hat{\mathbf{x}}_i(t_k), \mathbf{i}_i(t_k), \hat{\mathbf{i}}_i(t_k); \theta, \xi)$, $t_k$ is the measurement time instant, and $n$ is the total number of measurement points. The weighted MSE between predicted curves and ground-truth curves is adopted as the loss function:

$$\mathcal{L}(t_k) = \mathbf{w}_{\mathbf{x}}^T \left\| \mathbf{x}(t_k) - \hat{\mathbf{x}}(t_k) \right\|_2 + \mathbf{w}_{\mathbf{i}}^T \left\| \mathbf{i}(t_k) - \hat{\mathbf{i}}(t_k) \right\|_2 \quad (27)$$

where $\mathbf{w}$ denotes the weighting factors of different variables. Then, the SGD method or the Adam optimizer can be applied. The training procedures are demonstrated in Pseudo-code 1. For the autoencoder-based frameworks, the reconstruction losses of the autoencoders are also added to (27):

$$\mathcal{L}(t_k) = (\mathbf{w}_{\mathbf{x}})^T \left\| \mathbf{x}(t_k) - \hat{\mathbf{x}}(t_k) \right\|_2 + (\mathbf{w}_{\mathbf{i}})^T \left\| \mathbf{i}(t_k) - \hat{\mathbf{i}}(t_k) \right\|_2$$
$$+ (\mathbf{w}_{\mathbf{x}}^{AE})^T \left\| dec_{\mathbf{X}} \{ enc_{\mathbf{x}} [\mathbf{x}(t_k)] \} - \mathbf{x}(t_k) \right\|_2 \quad (28)$$
$$+ (\mathbf{w}_{\mathbf{i}}^{AE})^T \left\| dec_{\mathbf{I}} \{ enc_{\mathbf{i}} [\mathbf{i}(t_k)] \} - \mathbf{i}(t_k) \right\|_2$$

where $\mathbf{w}^{AE}$ denotes the weighting factors of reconstruction losses of different variables.

## IV. NEURAL DYNAMIC MODEL-INTEGRATED TRANSIENT STABILITY SIMULATIONS

### A. Basics of Power System Transient Stability Simulations

The system DAEs shown in (4) and (5) can be solved by the simultaneous approach or by the alternating approach. In the simultaneous approach, after applying implicit numerical integration methods to system ODEs, the ODEs will transform into algebraic equations and can be solved simultaneously with (5) using the Newton method iteratively, as shown in (29)-(30):

$$\begin{bmatrix} \mathbf{F}(\mathbf{x}, \mathbf{V}) \\ \mathbf{G}(\mathbf{x}, \mathbf{V}) \end{bmatrix} = \begin{bmatrix} \mathbf{0} \\ \mathbf{0} \end{bmatrix} \quad (29)$$

$$\begin{bmatrix} \dfrac{\partial \mathbf{F}}{\partial \mathbf{x}} & \dfrac{\partial \mathbf{F}}{\partial \mathbf{V}} \\ \dfrac{\partial \mathbf{G}}{\partial \mathbf{x}} & \dfrac{\partial \mathbf{G}}{\partial \mathbf{V}} \end{bmatrix} \begin{bmatrix} \Delta \mathbf{x} \\ \Delta \mathbf{V} \end{bmatrix} = \begin{bmatrix} -\mathbf{F} \\ -\mathbf{G} \end{bmatrix} \quad (30)$$

where $\mathbf{F}(\mathbf{x}, \mathbf{V})$ are the transformed algebraic equations of system ODEs. An example of the implicit trapezoidal method is shown in (31):

$$\mathbf{F}(\mathbf{x}, \mathbf{V}, t) = \mathbf{x}(t)$$
$$- \left\{ \mathbf{x}(t - \Delta t) + \dfrac{\Delta t}{2} \left[ \mathbf{f}(\mathbf{x}(t), \mathbf{V}(t)) + \mathbf{f}(\mathbf{x}(t - \Delta t), \mathbf{V}(t - \Delta t)) \right] \right\} \quad (31)$$

On the other hand, in the alternating approach, the ODEs and AEs are solved separately. Fast solution techniques [28] can be used to improve numerical stability and accelerate the solution processes. Firstly, given the state vector $\mathbf{x}(t)$ and bus voltage vector $\mathbf{V}(t)$ at instant $t$, estimate the initial value $\mathbf{V}^{(0)}(t + \Delta t)$, or simply let $\mathbf{V}^{(0)}(t + \Delta t) = \mathbf{V}(t)$. Secondly, solve (4) to obtain $\mathbf{x}^{(0)}(t + \Delta t)$ using the numerical integration method with $\mathbf{x} = \mathbf{x}(t)$ and $\mathbf{V} = \mathbf{V}^{(0)}(t + \Delta t)$. Third, solve (5) for $\mathbf{V}^{(1)}(t + \Delta t)$ with $\mathbf{x} = \mathbf{x}^{(0)}(t + \Delta t)$ and $\mathbf{V} = \mathbf{V}^{(0)}(t + \Delta t)$. Then solve (4) again with $\mathbf{x} = \mathbf{x}^{(0)}(t + \Delta t)$ and $\mathbf{V} = \mathbf{V}^{(1)}(t + \Delta t)$. The above iteration is repeated until convergence is reached.

As can be seen, the simultaneous approach is more rigorous and the dynamic models must support the calculation of partial derivatives shown in (30). In contrast, the alternating approach will cause errors, but this approach does not require partial derivatives and is widely used in industry-grade simulation programs because of its programming flexibility and simplicity, reliability, and robustness [23].

### B. Integrating Neural Models into Simulations

Firstly, the simulator should be able to load neural networks and perform forward propagation of neural networks.

As for the simultaneous approach, the simulator needs to additionally support the backward propagation of neural networks.

Taking a neural DAE model and the implicit trapezoidal method as an example, the algebraic integration function is shown in (32) and the injection currents are calculated as shown in (33). The partial derivatives with respect to the input vector of the neural module are needed, as shown in (34):

$$\mathbf{F}(\mathbf{x}, \mathbf{i}, \mathbf{v}, \mathbf{z}, t) = \mathbf{x}(t) - \left\{ \mathbf{x}(t - \Delta t) + \frac{1}{2} \Delta t \left[ \mathbf{\Psi}(t) + \mathbf{\Psi}(t - \Delta t) \right] \right\} \quad (32)$$

$$\mathbf{i}(t) = \boldsymbol{\varphi}(t) \quad (33)$$

$$\begin{bmatrix} \dfrac{\partial \mathbf{F}}{\partial \mathbf{x}} & \dfrac{\partial \mathbf{F}}{\partial \mathbf{v}} \\ \dfrac{\partial \mathbf{G}}{\partial \mathbf{x}} & \dfrac{\partial \mathbf{G}}{\partial \mathbf{v}} \end{bmatrix} = \begin{bmatrix} \mathbf{J}_{11}\left(\dfrac{\partial \mathbf{\Psi}}{\partial \mathbf{x}}\right) & \mathbf{J}_{12}\left(\dfrac{\partial \mathbf{\Psi}}{\partial \mathbf{v}}\right) \\ \mathbf{J}_{21}\left(\dfrac{\partial \boldsymbol{\varphi}}{\partial \mathbf{x}}\right) & \mathbf{J}_{22}\left(\dfrac{\partial \boldsymbol{\varphi}}{\partial \mathbf{v}}\right) \end{bmatrix} \quad (34)$$

where $\mathbf{J}(*)$ represents a function of $*$.

As for the alternating approach, the integration is intuitive. The neural ODE function $\mathbf{\Psi}$ is used as the derivative function. It can be easily solved by numerical integration methods. The neural AE function $\boldsymbol{\varphi}$ is used to calculate injection currents. After adding the calculated injection currents to the connected bus, nodal voltages can be obtained by solving network AEs (5).

To summarize, the neural models derived by the proposed neural ODE-E and DAE modules can be integrated into transient simulations and will not affect the simulation procedures.

*C. Initial Value Learner*

For a dynamic component, the initial values of state variables cannot be directly accessed, e.g., the initial value of the rotor angle is determined by the nodal voltage and power generation obtained from the power flow solution. It means that in simulations, the initial value $\mathbf{x}(0)$ is not available for the neural models. Therefore, an initial value learner is needed when integrating neural models into transient stability simulations. In most cases, the learner is of the form shown in (35):

$$\mathbf{x}(0) = \mathbf{h}(\mathbf{i}, \mathbf{v}, \mathbf{z}; \boldsymbol{\zeta}) \quad (35)$$

where $\mathbf{h}$ is the initial value calculator and $\boldsymbol{\zeta}$ denotes the parameters. The learner is also trained with the neural blocks.

*D. Discrete Event Handling*

During simulations, events such as faults and relay actions will introduce jump changes to the network and operation variables such as nodal voltages. Taking Fig. 2. (b) as an example, the state variable $\mathbf{x}(t)$ will not change right before and after the event, i.e., $\mathbf{x}(t^+) = \mathbf{x}(t^-)$, whereas some variables in $\mathbf{z}(t)$ can change, i.e., $\mathbf{z}(t^+) \neq \mathbf{z}(t^-)$. The difference between $\mathbf{z}(t^+)$ and $\mathbf{z}(t^-)$ could be large when the power system is subjected to a major failure. Considering these jump changes during the training process can improve the accuracy of the neural models. However, in most power system simulation tools, jump changes cannot be obtained. An acceptable alternative is to use $\mathbf{z}(t+1)$ instead of $\mathbf{z}(t)$ to predict $\dot{\mathbf{x}}(t)$ because the variables at the next integration step are a good approximation of the jump changes. Additional equality constraint (36) is added to (10) and constraints (36), (37), and (38) are added to (18):

$$\hat{\mathbf{z}}_i(t^+) = \hat{\mathbf{z}}_i(t+1), t \in \mathbf{T}_{event} \quad (36)$$

$$\hat{\mathbf{v}}_i(t^+) = \hat{\mathbf{v}}_i(t+1), t \in \mathbf{T}_{event} \quad (37)$$

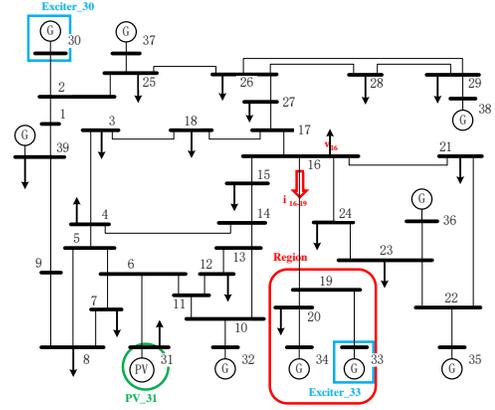

Fig. 3. The IEEE-39 test system and locations of components.

$$\mathbf{i}_i(t^+) = \boldsymbol{\varphi}(\mathbf{x}_i(t), \hat{\mathbf{v}}_i(t^+), \hat{\mathbf{z}}_i(t^+); \boldsymbol{\xi}), t \in \mathbf{T}_{event} \quad (38)$$

where $\mathbf{T}_{event}$ is the set of time instants when events happen. The jump change $\mathbf{i}_i(t^+)$ is calculated with $\boldsymbol{\varphi}(\mathbf{x}_i(t), \hat{\mathbf{v}}_i(t^+), \hat{\mathbf{z}}_i(t^+); \boldsymbol{\xi})$. Then the time derivative $\mathbf{\Psi}(\mathbf{x}_i(t), \mathbf{i}_i(t^+), \hat{\mathbf{v}}_i(t^+), \hat{\mathbf{z}}_i(t^+); \boldsymbol{\theta})$ is used to calculate $\mathbf{x}_i(t + \Delta t)$.

*E. Convergence Maintenance*

The DAE models trained by the neural DAE modules are connected to the power network as a simple current source of which the current injection can be calculated as shown in (17). However, a current source without a shunt admittance can seriously affect the convergence in most cases because the network equations may lose diagonal dominance. Therefore, besides the trained neural DAE model, an additional shunt admittance is also connected to the power network, and the injection currents will be recalculated as follows:

$$\mathbf{i}' = \mathbf{i} + \mathbf{v} \cdot (jB') \quad (39)$$

where $\mathbf{i}'$ is the fictitious injection current and $B'$ is the fictitious susceptance, which is set to $-50.0$ in this paper. Introducing the fictitious susceptance can maintain the diagonal dominance of the network equations without affecting accuracy [15].

## V. Module Designs for Typical Power Components

In this paper, three simple but representative cases are tested to demonstrate the validity and potentiality of the neural modules. The IEEE-39 system in Fig. 3 is used for model training and validation. Simulated variables are used as accessible measurement data. In the IEEE-39 system, the components are modeled in detail including the sixth-order generator model, exciters, governors and movers, power system stabilizers, and induction motor load models. A photovoltaic (PV) power plant is connected to bus 31 and the PV power plant model [29] of the Power System Analysis Software Package is adopted, which is developed by the China Electric Power Research Institute.

Two excitation controllers *Exciter_30* and *Exciter_33* marked in the blue rectangles, one PV power plant *PV_31* marked in the green circle, and a regional power network *Region* marked in the red rounded rectangle are used as representative components to test the feasibility of the neural ODE-E module and the neural DAE module. The trained models are integrated into a power system electromechanical simulator to test the performance of neural models in actual transient simulations. Finally, the neural models trained in the IEEE-39 system are directly integrated into the 2383wp system, which contains 2383 buses, 2892 branches, 327 generators, and 1822 loads, to test the performance of the neural models in large-scale power systems. The detailed information of the 2383wp system can be found in [30].

### A. Neural ODE-E Module Designs for Black-box Exciters

A common excitation controller model displayed in Fig. 4 is used as a representative vendor-specific controller to test the feasibility of the neural ODE-E module. Only the portal measurements, i.e., the input variables $V$ and $V_S$ and the output variable $E_{fd}$, of the exciter are accessible, i.e., the exciter is a black box for simulations. A data-driven model with knowable structure and parameters can be obtained using the neural ODE-E module.

Two exciters with different parameters are tested. The $E_{fd\max}$ and $E_{fd\min}$ of the Exciter_30 model at bus 30 are respectively set to plus and minus infinity to test the generalization ability of the neural ODE-E module, whereas those of Exciter_33 at bus 33 are respectively set to 0.0 and 3.3 to test the neural ODE-E module's ability to handle nonlinearity. The loss function and the interface settings are shown in (40):

$$\begin{aligned} \mathcal{L}(t_k) &= \|\mathbf{x}(t_k) - \hat{\mathbf{x}}(t_k)\|_2 \\ \mathbf{x}(t) &= \{E_{fd}(t)\} \\ \mathbf{z}(t) &= \{V(t), V_S(t), V(0), V_S(0), \mathbf{x}(0)\} \end{aligned} \quad (40)$$

In $\mathbf{z}$, the initial steady-state values of $V$, $V_S$, and $E_{fd}$ are included so that the neural ODE block can learn how to calculate the reference variables $V_{\text{ref}}$ and $E_{fd\text{ref}}$. In most cases, $V_{\text{ref}}$ and $E_{fd\text{ref}}$ are unknown and are calculated based on the steady state of the exciter. If the reference variables are known, they can be directly used to form $\mathbf{z}$.

### B. Neural DAE Module Designs for a PV Power Plant

A PV power plant model [29] is used to test the feasibility of the neural DAE module for modeling renewables. Since the state variables are hard to obtain in practical power systems, the models are trained based on the portal measurements, i.e., the nodal voltage $\mathbf{v}$ and the injection $\mathbf{i}$.

With only the portal measurements, the loss function, the interface settings, and the initial value learner are shown in (41):

$$\begin{aligned} \mathcal{L}(t_k) &= \|\mathbf{i}(t_k) - \hat{\mathbf{i}}(t_k)\|_2 \\ \begin{cases} \mathbf{x}(t) = \tilde{\mathbf{x}}(t) \\ \mathbf{i}(t) = \{i_x(t), i_y(t)\} \\ \mathbf{v}(t) = \{v_x(t), v_y(t)\} \\ \mathbf{z}(t) = \{v_x(0), v_y(0), i_x(0), i_y(0)\} \end{cases} \\ \mathbf{x}(0) &= \tilde{\mathbf{x}}(0) = \mathbf{h}[\mathbf{v}(0), \mathbf{i}(0); \zeta] \end{aligned} \quad (41)$$

where $\tilde{\mathbf{x}}$ is the fictitious state variable vector, which is calculated with an initial value learner $\mathbf{h}$, $v_x$ and $v_y$ are respectively the real part and imaginary part of the nodal voltage of bus 31, and $i_x$ and $i_y$ are respectively the real part and imaginary part of the injection current. An MLP $\mathbf{h}$ with two hidden layers that learns the initial values $\mathbf{x}(0)$ is established. The trained PV model can be directly integrated into transient simulations.

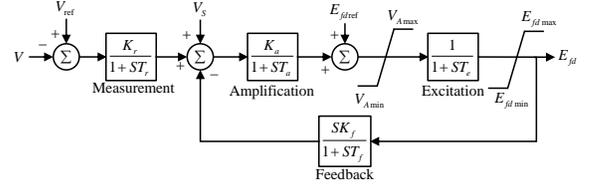

Fig. 4. Excitation system model. The input $V$ is the amplitude of the nodal voltage. The model contains four blocks including measurement, amplification, excitation, and feedback. $K$ and $T$ denote the enlargement factor and the time constant of each block, respectively. $V_S$ represents the additional input signal from the power system stabilizer. $V_{\text{ref}}$ and $E_{fd\text{ref}}$ are the reference variables.

It should be noted that the fictitious state vector $\tilde{\mathbf{x}}$ is designed because the actual state variables are not accessible. If there are accessible state variables, e.g., the DC voltage and current of the PV cells, these state variables can be directly used as $\mathbf{x}$ or as part of $\mathbf{x}$. Meanwhile, the accessible state variables need to be considered in the loss function as (40).

### C. Neural DAE Module Designs for an Equivalent Load

A regional power network that consists of buses 19, 20, 33, and 34 is used as a composite load to test the feasibility of the proposed neural DAE module for equivalent load modeling. This is a typical situation when building an equivalent load for distribution networks, microgrids, etc. The generators and the load in this region are modeled in detail, i.e., the internal dynamics of the power network are complex. Bus 16 is similar to a point of common coupling. Similarly, all the state variables are not accessible except for the portal measurements, i.e., the nodal voltage of bus 16 $\mathbf{v}_{16}$ and the transmission current from bus 16 to bus 19 $\mathbf{i}_{16-19}$. An equivalent load model of the regional power network can be derived by the neural DAE module.

Similarly, with the portal measurements, the loss function, the interface settings, and the initial value learner are also shown in (41), where $v_x$ and $v_y$ are respectively the real part and imaginary part of $\mathbf{v}_{16}$, and $i_x$ and $i_y$ are respectively the real part and imaginary part of $\mathbf{i}_{16-19}$. The trained equivalent load model can be directly integrated into simulations.

### D. Building Neural Modules with Neural Networks

For the regular neural ODE-E and DAE modules in Fig. 2. (b) and Fig. 2. (d), the neural ODE block and the neural AE block are built using MLP with three hidden layers. All the hidden layers have the same width, which is defined as the dimensionality of the hidden layers. For the autoencoder-based neural ODE-E and DAE modules in Fig. 2. (c) and Fig. 2. (e), the encoder, decoder, the neural ODE block, and the neural AE block are all constructed using MLP with one hidden layer. All the hidden layers also have the same width, which is defined as the dimensionality of the hidden space. Taking the neural ODE-E module as an example, after encoding, $\mathbf{X}$ and $\mathbf{Z}$ will have the same dimensionality as the dimensionality of the hidden space.

In this paper, between any two adjacent neural layers, the ex-ponential linear unit (ELU) [31] is used as the activation function for its good performance. There are no activation functions for output layers since the outputs of derivative functions are not restricted. A gradient clipping technique is adopted to handle the problem of exploding gradient.

## VI. IMPLEMENTATION AND NUMERICAL TESTS

### A. Implementation Overview

The test platform is a Linux server consisting of one Intel i7-10700KF 3.80GHz 8-core CPU, one NVIDIA GeForce RTX 3090 GPU, and 128GB DDR4-3200MHz RAM.

The neural modules are developed with Python based on an open-source neural ODE package called torchdiffeq on GitHub [9]. The source code of the proposed neural ODE-E module and neural DAE module is made public on GitHub. An alternating approach-based high-performance transient simulator, which is purely written in C++ based on our previous studies [30], [32], and [33] and realizes neural network supportability using the PyTorch C++ application programming interface (API) called LibTorch, is adopted to generate sample curves and test trained neural models. The Python API of the simulator is also shared on GitHub [34]. The simulator can directly load the structure and parameters of neural networks and perform forward and backward propagations of neural networks.

Three ODE solvers based on the Euler method, the midpoint method, and the fourth-order Runge-Kutta (RK4) method, are implemented in the source code of the proposed neural modules. On the other hand, in the power system simulator, the implicit trapezoidal method is utilized because it is numerically A-stable [23]. The simulation time is 10 seconds and the integration step is 0.01 seconds.

### B. Training and Testing Designs

#### 1) Dataset Designs

The test system used for model training is the standard IEEE-39 system as shown in Fig. 3. Power flow scenarios are sampled by randomly changing the states of generators and loads. The nodal voltage of a generator is sampled between 0.94 p.u. and 1.06 p.u. The active power generation of a generator ranges from its lower generation limit to its upper generation limit. The active power and the reactive power of a load are randomly sampled between 50 percent and 150 percent of the load power given in the standard IEEE-39 system. Each sampled operating state is subject to a random fault or disturbance including three-phase short-circuit faults, generator tripping, load shedding, etc.

For each representative component, a total number of 4000 samples are generated, of which 800 samples are used as the testing set and the rest 3200 samples are used as the training set. As a feasibility study, two different datasets are generated. One dataset (hereafter referred to as *Dataset_A*) is generated to ensure the diversity of samples, in which half of the samples maintain rotor angle stability whereas the other half of the samples lose rotor angle stability. It should be noted that only the state variables and algebraic variables before the maximum rotor angle difference exceeds 360 degrees are used to train the neural models. The other dataset (hereafter referred to as *Dataset_B*) is generated under a more practical consideration, in which all

TABLE I
TRAINING SETTINGS FOR DIFFERENT COMPONENTS

| Component | Framework | $N_r$ | $n_d$ | $E$ | $l_r$ | $\gamma$ |
|---|---|---|---|---|---|---|
| Exciter_30 without limit | Regular | 200 | 64 | 50 | | |
| | Autoencoder | 200 | 16 | 400 | | |
| Exciter_33 with limit | Regular | 400 | 64 | 200 | | |
| | Autoencoder | 200 | 32 | 200 | 0.005 | 0.7 |
| PV_31 | Regular | 400 | 64 | 200 | | |
| | Autoencoder | 400 | 64 | 200 | | |
| Region | Regular | 800 | 64 | 400 | | |
| | Autoencoder | 800 | 64 | 400 | | |

In the table, $N_r$, $n_d$, $E$, $l_r$, and $\gamma$ are the size of the training set, the dimensionality of the hidden layers or the hidden space, the number of training epochs, the learning rate, and the learning rate damping factor, respectively.

the samples are stable. Meanwhile, only 20 percent of the samples consider three-phase short-circuit faults and the remaining 80 percent of samples consider generation changes and load changes. The generation and load changes range from 10 percent and 90 percent and are randomly selected. Therefore, with both datasets, the neural models are trained to capture the dynamics within the stability boundary.

#### 2) Training Setting Designs

The Euler-based ODE solver and the Adam optimizer are adopted to train neural models. Tests are performed with different training settings to check the requirements and performance of each module. In the tested settings, the size of the training set $N_r$ includes 3200, 1600, 800, 400, 200, 100, and 50, the dimensionality of hidden layers or the hidden space $n_d$ includes 64, 32, 16, and 8, the number of training epochs $E$ includes 400, 200, 100, and 50, the learning rate $l_r$ includes 0.005, and 0.001, and the learning rate damping factor $\gamma$ includes 0.7 and 0.5. A few tests with the largest $N_r$, $n_d$, $E$ are performed to check the influence of learning rates. With all the learning rate settings mentioned before, acceptable models can be obtained. Therefore, $l_r$ and $\gamma$ are fixed at 0.005 and 0.7 respectively.

For each module, the training tests are carried out in the following three steps. Firstly, decrease the size of the training set to find out how many samples are required for obtaining a model with acceptable performance. Secondly, decrease the complexity of neural modules, which is the width of neural layers in this paper. At last, decrease the number of training epochs to accelerate the training process. After tests, the training settings adopted for different components are summarized in Table I.

#### 3) Test Design of Neural Model-Integrated Simulation

After training, each neural model is integrated into the simulator and is tested under 800 new scenarios with randomly sampled operating states and contingencies. There are 698 stable scenarios and 102 unstable scenarios. As mentioned before, the neural models are trained to approximate the dynamics within the stability boundary. Therefore, only the state variables and algebraic variables of the original test system and the neural model-integrated test system before the systems lose stability, as well as the time instant when the system systems lose stability are compared to verify the effectiveness of the proposed neural modules. The comparative simulation results of the original model-integrated simulations and the neural model-integrated

TABLE II
COMPARATIVE SIMULATION RESULTS UNDER 800 NEW SCENARIOS OF DIFFERENT MODELS TRAINED BY THE EULER METHOD

| Component and key variable | Framework | Dataset_A $\Delta x$ | Dataset_A $\Delta T_s$ (s) | Dataset_B $\Delta x$ | Dataset_B $\Delta T_s$ (s) |
|---|---|---|---|---|---|
| Exciter_30, $E_{fd}$ (p.u.) | Regular | 1.86E-01 | 4.71E-03 | 2.27E-01 | 4.90E-03 |
| | Autoencoder | 3.81E-02 | 5.88E-04 | 4.94E-02 | 1.08E-03 |
| Exciter_33, $E_{fd}$ (p.u.) | Regular | 1.28E-02 | 8.82E-04 | 1.56E-02 | 7.84E-04 |
| | Autoencoder | 7.37E-03 | 3.92E-04 | 7.84E-03 | 4.90E-04 |
| PV_31, $P_{pv}$ (p.u.) | Regular | 1.76E-01 | 1.78E-02 | 1.57E-01 | 2.01E-02 |
| | Autoencoder | 1.54E-01 | 1.51E-02 | 1.43E-01 | 1.75E-02 |
| Region, $P\_L$ (p.u.) | Regular | 3.24E-01 | 5.57E-02 | 3.75E-01 | 7.09E-02 |
| | Autoencoder | 2.73E-01 | 4.12E-02 | 3.27E-01 | 5.84E-02 |

In the table, $\Delta x$ denotes the average absolute modeling errors of the key variables per step and $\Delta T_s$ denotes the average deviation of the time instant when the system loses stability per sample.

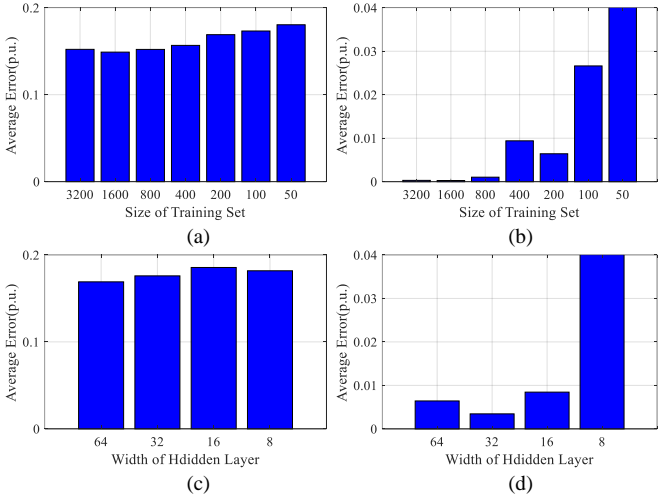

Fig. 5. Average errors of the neural ODE-E module for the Exciter_30 model in the test set of 800 samples. The left column belongs to the regular structure and the right column belongs to the autoencoder-based structure.

simulations of all four dynamic components are summarized in Table II. Detailed test results are as follows.

### C. Testing Details of the Excitation Controllers

#### 1) Tests of Different Training Settings

To avoid redundancy, only the results of the three-step training test of the Exciter_30 model with Dataset_A are illustrated. In Fig. 5, the average errors of the neural ODE-E module for the Exciter_30 model in the test set are displayed. In Fig. 5. (a) and Fig. 5. (b), the sample requirements of the proposed modules are tested. In Fig. 5. (c) and Fig. 5. (d), the neural module performance with different $n_d$ settings is displayed. Similar results can be obtained with Dataset_B and different components including the Exciter_33 model, the PV_31 model, and the Region model.

As can be seen from Table I, with hundreds of samples, which is not very difficult to obtain in practical power systems, both modules obtained acceptable models. In particular, the autoencoder-based neural ODE module can provide highly accurate approximations of the original dynamics of the component.

#### 2) Accuracy of Neural Model-Integrated Simulations

The trained neural models are integrated into the power system simulator. Comparative simulation results are displayed in

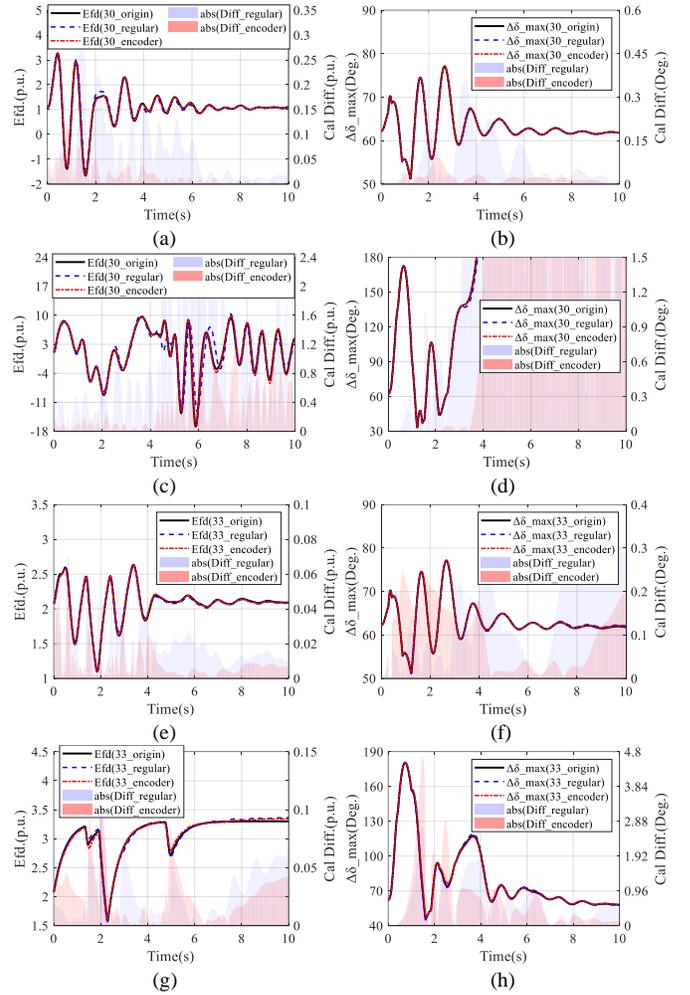

Fig. 6. Comparative results of the original model-integrated simulations and the neural excitation controller model-integrated simulations.

Table II and Fig. 6. In Fig. 6 and the following figures showing the comparative results of different models, "origin" denotes the original model, "regular" denotes the neural model without autoencoder, "encoder" denotes the autoencoder-based neural model, "$\Delta \delta\_\max$" denotes the system-wide maximum rotor angle difference, and "abs(Diff_*)" denotes the absolute errors between the original model and the model represented by "*".

For the Exciter_30 model, when the system is stable as in Figs. 6. (a) and 6. (b), and before the time instant 3.72 seconds in Figs. 6. (c), 6. (d), the neural models perform well, whereas when the system loses stability as in Figs. 6. (c), 6. (d), after 3.72 seconds, the errors between the learned dynamics and the ground-truth dynamics increase. This phenomenon is consistent with the training designs because only the dynamics in the stability region are used to train the neural models. As long as the simulation gives an unstable prediction for an unstable contingency and the deviation of the time instant when the system loses stability remains in an acceptable range, the neural models can be used. As in Figs. 6. (e), 6. (f), 6. (g), 6. (h), the output limit of the Exciter_33 model is captured by the neural model, which proves the neural ODE-E module's ability to handle nonlinearity due to the nonlinear ELU activation.

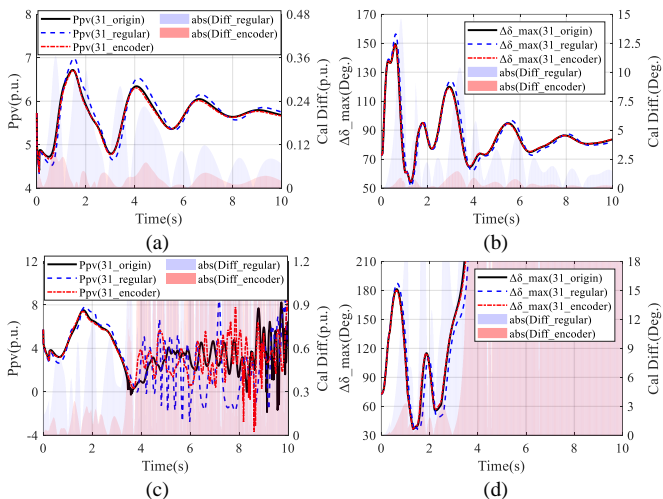

Fig. 7. Comparative results of the original model-integrated simulations and the neural PV model-integrated simulations.

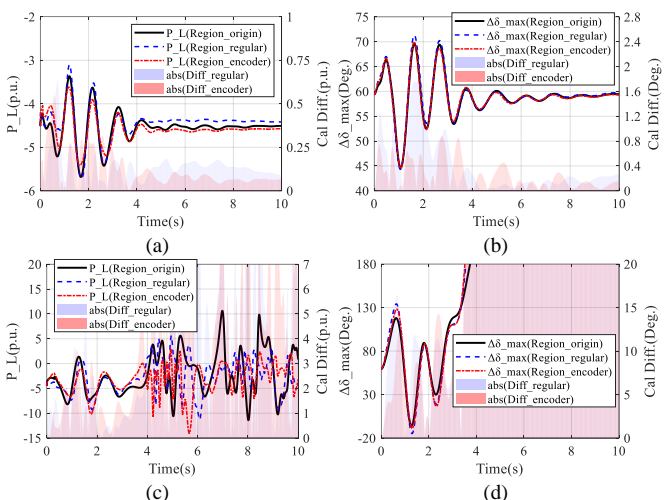

Fig. 8. Comparative results of the original model-integrated simulations and the neural equivalent load model-integrated simulations.

In Table II, the autoencoder-based neural exciter models outperform the regular neural exciter models with smaller $E_{fd}$ errors and smaller error propagation through the power network. On the other hand, with Dataset_B that only contains stable samples, feasible neural models can still be obtained, and the performance of the neural models just degrades slightly. The time deviation of instability moment of the models that are trained with Dataset_B becomes larger. This is intuitive because the dynamics near the stability boundary are not well learned due to the lack of unstable samples.

### D. Testing Details of the PV Power Plant

As shown in Table I, four hundred samples are used to train data-driven models of the PV_31 model. With both datasets, acceptable neural models are obtained with the training settings. The trained models are integrated into the simulator. Comparative simulation results can be found in Table II and Fig. 7, where $P_{pv}$ denotes the active power output of the PV power plant. Compared with the exciter models, the neural PV models introduce more errors to the whole power network. The reason is intuitive. For a controller such as an exciter, the modeling error propagates to the whole network by affecting the state variables of the controlled device such as a generator. In contrast, for a power device that directly injects or draws power from the power network such as a generator, a PV power plant, etc., the modeling error directly propagates to the power network through the injection current. The average deviation of the time instant when the system loses stability increases with the modeling errors. The simulations provide correct predictions of system stability, and the time deviation of instability moment is within an acceptable range.

### E. Testing Details of the Equivalent Load Modeling

As displayed in Table I, eight hundred samples are utilized. With both datasets, the adopted training settings in Table I obtains acceptable neural models. Compared with the exciter and the PV model, more samples are required to train an equivalent load model for a regional power network because the dynamics of the inner components become much more complex and the accessible data only contains portal measurements, i.e., the algebraic variables $\mathbf{v}_{16}$ and $\mathbf{i}_{16-19}$.

The trained models are integrated into the simulator. Comparative simulation results are shown in Table II and Fig. 8, where $P\_L$ denotes equivalent load power and the active transmission power from bus 16 to bus 19 in the original network. The test results indicate that the neural DAE module can be used to obtain an equivalent load model for a composite load such as a regional power network by learning the hidden dynamics based only on the algebraic variables.

### F. Integration Method Tests

The RK4-based ODE solver is also used to train neural models with Dataset_B. The simulation results of the neural models trained with the Euler method and the RK4 method are displayed in Table III, where the minimum values of $\Delta x$ and $\Delta T_s$ are bolded. It can be seen that although different integration methods are adopted in the training process, after being integrated into the traditional transient simulation with the implicit trapezoidal method, the obtained neural models all achieve acceptable accuracy. The RK4 method tends to lead to more precise models. On the other hand, the time consumption of the RK4 method is about 6 times that of the Euler method.

### G. Simulation Tests in 2383wp System

The neural models of two exciters, the PV power plant, and the equivalent load model, which are trained in the IEEE-39 system with Dataset_B and the Euler method, are directly integrated into the 2383wp system. The exciters of generators at buses 10 and 18 of the 2383wp system are replaced with the Exciter_30 and Exciter_33 models. The generator at bus 16 of the 2383wp system is replaced with the PV_31 model. The buses 19, 20, 33, and 34 of the IEEE-39 system are connected to bus 322 of the 2383wp system, i.e., the modified 2383wp system contains 2387 buses. 100 scenarios are tested. In the tested scenarios, 37 scenarios are unstable. The simulation results are displayed in Table IV.

TABLE III
COMPARATIVE SIMULATION RESULTS OF DIFFERENT INTEGRATION METHODS

| Component and key variable | Framework | Dataset_B | | | |
|---|---|---|---|---|---|
| | | Euler | | RK4 | |
| | | $\Delta x$ | $\Delta T_s$ (s) | $\Delta x$ | $\Delta T_s$ (s) |
| Exciter_30, $E_{fd}$ (p.u.) | Regular | 2.27E-01 | 4.90E-03 | **1.96E-01** | **4.02E-03** |
| | Autoencoder | **4.94E-02** | 1.08E-03 | 5.78E-02 | **6.86E-04** |
| Exciter_33, $E_{fd}$ (p.u.) | Regular | 1.56E-02 | **7.84E-04** | **1.52E-02** | 8.82E-04 |
| | Autoencoder | **7.84E-03** | 4.90E-04 | 8.28E-03 | **3.92E-04** |
| PV_31, $P_{pv}$ (p.u.) | Regular | 1.57E-01 | 2.01E-02 | **1.54E-01** | **1.89E-02** |
| | Autoencoder | **1.43E-01** | 1.75E-02 | 1.44E-01 | **1.67E-02** |
| Region, $P\_L$ (p.u.) | Regular | 3.75E-01 | 7.09E-02 | **3.63E-01** | **6.25E-02** |
| | Autoencoder | **3.27E-01** | 5.84E-02 | **3.25E-01** | 4.50E-02 |

TABLE IV
COMPARATIVE SIMULATION RESULTS OF DIFFERENT MODELS UNDER 800 NEW SCENARIOS IN THE 2383WP SYSTEM

| Component and key variable | Framework | Dataset_B | |
|---|---|---|---|
| | | $\Delta x$ | $\Delta T_s$ (s) |
| Exciter_30, $E_{fd}$ (p.u.) | Regular | 2.40E-02 | 0.00 |
| | Autoencoder | 6.20E-03 | 0.00 |
| Exciter_33, $E_{fd}$ (p.u.) | Regular | 6.69E-03 | 0.00 |
| | Autoencoder | 2.59E-03 | 0.00 |
| PV_31, $P_{pv}$ (p.u.) | Regular | 5.80E-02 | 0.00 |
| | Autoencoder | 3.27E-02 | 0.00 |
| Region, $P\_L$ (p.u.) | Regular | 1.49E-01 | 0.00 |
| | Autoencoder | 1.12E-01 | 0.00 |

Although the neural models are only trained by samples of the IEEE-39 system, these neural models work properly in the modified 2383wp system. The modeling errors remain small. The average time deviation of instability is equal to 0.0, i.e., the stability predictions are not affected at all. The influence of the modeling error of a single component is limited due to the large electrical distance.

## VII. DISCUSSIONS AND FUTURE WORKING DIRECTIONS

### A. Learning Dynamics Beyond Stability Boundary

As mentioned before, the neural models are trained to capture the component dynamics in the stability region. Unstable system dynamics are required for training a data-driven model that can generalize beyond the stability boundary. However, it is hard to acquire credible unstable dynamics because unstable dynamics can only be simulated but the integration errors accumulate large rapidly when the system is unstable. Therefore, in the numerical tests, the learned dynamics are theoretically credible only in the stability region. The generalization of the system dynamics beyond the stability boundary remains a challenging issue for data-driven models. Hybrid modeling methods based on the proposed neural ODE-E and DAE modules are a future working direction to address this issue.

### B. Adaptation to Electromagnetic Simulations

As for the accurate simulation of renewables, electromagnetic simulations are needed. The neural ODE-E module can be utilized in this case. Since the instantaneous values are calculated in the electromagnetic simulations, time $t$ may be considered in the component' ODEs, as in [35]. The uncertainty of component dynamics can be modeled with neural stochastic differential equations [36], which can be realized based on the proposed neural modules.

## VIII. CONCLUSION

In this paper, a neural ODE-E module and a neural DAE module for power system dynamic component modeling are proposed, which can build data-driven dynamic models based on accessible measurement data. The dynamic models trained by the proposed neural modules are directly integrated with analytical models using unified transient stability simulation methods to perform simulations simultaneously while maintaining the accuracy and convergence of solutions. Comparative results of the original model-based simulations and neural model-integrated simulations in the IEEE-39 system and the 2383wp system prove the feasibility of the proposed neural ODE-E and DAE modules. Test results indicate that the proposed neural modules can build accurate dynamic models for complex components based on a dataset of portal measurements that only contains stable samples with a limited proportion of large-disturbance contingencies. The source code of the proposed neural modules has been made public on GitHub.

## APPENDIX

### A. Proof of the Gradient Formulas

Given (21)-(23) and (26), the proof of the gradient formulas shown in (19) and (20) is as follows. Since $\mathbf{x}$ and $\mathbf{i}$ is determined by $\boldsymbol{\theta}$ and $\boldsymbol{\xi}$, $\mathcal{H} - \boldsymbol{\lambda}^T \boldsymbol{\Psi} - \boldsymbol{\beta}^T \boldsymbol{\Phi} = 0$, and (26), $\mathcal{L}$ can be reformulated as:

$$\mathcal{L} = \mathcal{L}(\boldsymbol{\theta}, \boldsymbol{\xi})$$
$$= \sum_{k=0}^{n} \mathcal{L}(t_k) + \int_0^T \left\{ \mathcal{H}(t) - [\boldsymbol{\lambda}(t)]^T \boldsymbol{\Psi}(t) - [\boldsymbol{\beta}(t)]^T \boldsymbol{\Phi}(t) \right\} dt \quad (42)$$

Differentiate $\mathcal{L}$ and (43) can be obtained.

$$\Delta \mathcal{L} \doteq \frac{\partial \mathcal{L}}{\partial \boldsymbol{\theta}} \Delta \boldsymbol{\theta} + \frac{\partial \mathcal{L}}{\partial \boldsymbol{\xi}} \Delta \boldsymbol{\xi} = \sum_{k=0}^{n} \Delta \mathcal{L}(t_k)$$
$$+ \int_0^T \left\{ \Delta \mathcal{H}(t) - \Delta \left\{ [\boldsymbol{\lambda}(t)]^T \boldsymbol{\Psi}(t) \right\} - \Delta \left\{ [\boldsymbol{\beta}(t)]^T \boldsymbol{\Phi}(t) \right\} \right\} dt \quad (43)$$

Since $\mathbf{i} = \boldsymbol{\varphi}$, each term in (43) can be differentiated as:

$$\Delta \mathcal{L}(t_k) \doteq \left[ \frac{\partial \mathcal{L}}{\partial \mathbf{x}(t_k)} + \frac{\partial [\boldsymbol{\varphi}(t_k)]^T}{\partial \mathbf{x}(t_k)} \frac{\partial \mathcal{L}}{\partial \mathbf{i}(t_k)} \right]^T \Delta \mathbf{x}(t_k) \quad (44)$$

$$\Delta \mathcal{H}(t) = -[\dot{\boldsymbol{\lambda}}(t)]^T \Delta \mathbf{x}(t) + \left[ \frac{\partial \mathcal{H}}{\partial \mathbf{i}} \right]^T \cdot \Delta \mathbf{i}(t) + \left[ \frac{\partial \mathcal{H}(t)}{\partial \boldsymbol{\theta}} \right]^T \Delta \boldsymbol{\theta}$$
$$+ \left[ \frac{\partial \mathcal{H}(t)}{\partial \boldsymbol{\lambda}} \right]^T \Delta \boldsymbol{\lambda}(t) + \left[ \frac{\partial \mathcal{H}(t)}{\partial \boldsymbol{\xi}} \right]^T \Delta \boldsymbol{\xi} + \left[ \frac{\partial \mathcal{H}(t)}{\partial \boldsymbol{\beta}} \right]^T \Delta \boldsymbol{\beta}(t) \quad (45)$$

$$\Delta \left\{ [\boldsymbol{\lambda}(t)]^T \boldsymbol{\Psi}(t) \right\} = [\boldsymbol{\lambda}(t)]^T \Delta \boldsymbol{\Psi}(t) + [\boldsymbol{\Psi}(t)]^T \Delta \boldsymbol{\lambda}(t) \quad (46)$$

$$\Delta \left\{ [\boldsymbol{\beta}(t)]^T \boldsymbol{\Phi}(t) \right\} = [\boldsymbol{\beta}(t)]^T \Delta \boldsymbol{\Phi}(t) + [\boldsymbol{\Phi}(t)]^T \Delta \boldsymbol{\beta}(t) \quad (47)$$

Substitute (44)-(47) to (43) and (48) can be derived.

$$\Delta \mathcal{L} = \frac{\partial \mathcal{L}}{\partial \boldsymbol{\theta}} \Delta \boldsymbol{\theta} + \frac{\partial \mathcal{L}}{\partial \boldsymbol{\xi}} \Delta \boldsymbol{\xi} = \sum_{k=0}^{n} \Delta \mathcal{L}(t_k) - \int_0^T d\left\{ [\boldsymbol{\lambda}(t)]^T \Delta \mathbf{x}(t) \right\}$$
$$+ \int_0^T \left\{ \left[ \frac{\partial \mathcal{H}(t)}{\partial \boldsymbol{\theta}} \right]^T \Delta \boldsymbol{\theta} + \left[ \frac{\partial \mathcal{H}(t)}{\partial \boldsymbol{\xi}} \right]^T \Delta \boldsymbol{\xi} \right\} dt \quad (48)$$

Let $\psi(t) = [\lambda(t)]^T \Delta \mathbf{x}(t)$. With (23), (49) can be derived.

$$\int_0^T d\psi(t) = \int_{t_0}^{t_n} d\psi(t) = \int_{t_0}^{t_1} d\psi(t) + \int_{t_1}^{t_2} d\psi(t) + \ldots + \int_{t_{n-1}}^{t_n} d\psi(t)$$
$$= \sum_{k=0}^n \left[ \frac{\partial \mathcal{L}}{\partial \mathbf{x}(t_k)} + \frac{\partial [\varphi(t_k)]^T}{\partial \mathbf{x}(t_k)} \frac{\partial \mathcal{L}}{\partial \mathbf{i}(t_k)} \right]^T \Delta \mathbf{x}(t_k) = \sum_{k=0}^n \Delta \mathcal{L}(t_k) \quad (49)$$

Finally, substitute (49) to (48) and derive:

$$\frac{\partial \mathcal{L}}{\partial \boldsymbol{\theta}} \Delta \boldsymbol{\theta} + \frac{\partial \mathcal{L}}{\partial \boldsymbol{\xi}} \Delta \boldsymbol{\xi} = \int_0^T \left\{ \left[ \frac{\partial \mathcal{H}(t)}{\partial \boldsymbol{\theta}} \right]^T \Delta \boldsymbol{\theta} + \left[ \frac{\partial \mathcal{H}(t)}{\partial \boldsymbol{\xi}} \right]^T \Delta \boldsymbol{\xi} \right\} dt \quad (50)$$

Since $\Delta \boldsymbol{\theta}$ and $\Delta \boldsymbol{\xi}$ are arbitrarily selected, the gradient formulas (19)-(20) are obtained. As for the gradient formulas shown in (11)-(14), the proof can be easily derived by removing the terms that contain $\boldsymbol{\Phi}$, $\mathbf{i}$, $\varphi$, and $\boldsymbol{\xi}$ in the above proof.